\documentclass[showpacs,twocolumn]{revtex4}

\usepackage{graphicx}
\usepackage{dcolumn}
\usepackage{amsmath}

\makeatletter
\def\btt#1{\texttt{\@backslashchar#1}}
\DeclareRobustCommand\bblash{\btt{\@backslashchar}} \makeatother

\begin{document}

\title{The role of the equation of state and the space-time dimension
in spherical collapse}

\author{Naresh Dadhich}
\email{nkd@iucaa.ernet.in} \affiliation{Inter-University Center
for Astronomy and Astrophysics, \\
 Post Bag 4, Ganeshkhind, Pune - 411 007, INDIA}%

\author{S. G.~Ghosh}
\email{sgghosh@iucaa.ernet.in}
\author{D. W.~Deshkar}
\affiliation{Department of Mathematics, Science College,
Congress Nagar, Nagpur 440 012, India}%

\date{\today}

\begin{abstract}
We study the spherically symmetric collapse of a fluid with
non-vanishing radial pressure in higher dimensional space-time. We
obtain the general exact solution in the closed form for the
equation of state ($P_r = \gamma \rho$) which leads to the
explicit construction of the root equation governing the nature
(black hole versus naked singularity) of the central singularity.
A remarkable feature of the root equation is its invariance for
the three cases: (${D+1},\; {\gamma = -1}$),
 (${D},\; {\gamma = 0}$) and (${D - 1}, \; {\gamma = 1}$) where $D$ is
the dimension of space-time. That is, for the ultimate end result
of the collapse, $D$-dimensional dust, ${D+1}$ - AdS (anti de
Sitter)-like and ${D-1}$ - dS-like are absolutely equivalent.
\end{abstract}


\maketitle

The tussle between black hole and naked singularity as the
ultimate end product of gravitational collapse is one of the
outstanding problems of classical general relativity (GR). In
spite of vigorous activity over two decades, we are far from
answering the question in a satisfactory manner. In fact we have
no more than a few conjectures, such as Penrose's  cosmic
censorship conjecture (CCC) \cite{rp} (see \cite{r1,rm} for
reviews on the CCC) and Thorne's hoop conjecture \cite{kt}, to go
by. On the other hand gravitational collapse under fairly general
conditions leads to singularity is very well established, thanks
to the elegant and powerful singularity theorems of Penrose and
Hawking \cite{he}.

The next important question is, whether or not singularity so
formed will causally influence any regular part of the space-time.
The CCC essentially says that a naked singularity (NS) which is
formed by evolution of {\it regular initial data} will be
completely shielded from the external view by an event horizon.
Such a singularity can be visible only to observers who fall
through the event horizon into the black hole (BH). That means
light rays can emanate from singularity but are completely blocked
by the event horizon and hence they could only lay bare to
observers who are co-falling with the collapsing star and never to
external observers. This is the weak CCC, while the strong CCC
prohibits its visibility by any observer. That means no light rays
emanate out of singularity, i.e., it is never naked. In the
precise mathematical terms it demands that space-time be globally
hyperbolic (for a given initial data, the dynamical evolution is
uniquely predictable). Existence of NS  would
therefore mean failure of global hyperbolicity and thereby
deterministic dynamics. This is why CCC is an essential ingredient
in a number of key theorems in GR, such as the black hole area
and the uniqueness theorems, and the positivity of mass theorem.

Of the two versions, the weak CCC seems to hold ground while there
do exist certain counter examples seriously challenging the strong
CCC \cite{rm}. It has been shown that it is possible to develop
NS from regular initial data. The simplest
setting for this is the spherical dust collapse described by the
Tolman-Bondi metric, which has been extensively studied \cite{dc}.
We have gained good bit of insight into the formation, visibility
and causal structure of the dust collapse singularities. All these
works neglect pressure which may play non trivial role in the
final outcome of the collapse. From this perspective,
gravitational collapse of perfect fluid has been studied \cite{pf}
to understand the role of pressure and the equation of state. It
however turns out that the presence of pressure does not
qualitatively alter the final outcome. In particular, the case of
radial pressure with vanishing tangential pressure has been
analyzed by Gon\c{c}alves and Jhingan \cite{gj} and it has been
shown, for an equation of state $P_r=\gamma\rho$, that the effect
of radial pressure for non negative $\gamma$ is to shrink the
parameter window in the initial data space giving rise to naked
singularity.  However, it could not prevent the formation of naked
singularity. On the other hand, when $\gamma = -1$, it is always
NS, completely violating CCC.

 In recent years, the string theory has
provoked explosive interest among theoretical physicists in
studying physics in higher dimensions (HD) \cite{rs}. While
gravitational collapse has been originally studied in four
dimensions (4D), there have been several attempts to study it in
HD space-time \cite{hvc,hdc,htt,hdo,hos,gj1}. Interestingly it turns out that
as dimension increases the parameter window for naked singularity
shrinks continuously. Recently, it has been conjectured that for a
marginally bound dust collapse, with initial density profile
sufficiently differentiable or smooth ($\rho_1=0$), the CCC is
always respected in HD with $D \ge 6$ \cite{gj1}. This is however
not true for the profile with $\rho_1$ non vanishing, where the
increase in $D$ only results in shrinking of the parameter window
leading to NS and in the non-marginally bound case even the
condition $\rho_1=0$ does not save CCC.

 In this study we would like to examine the role of radial pressure and
extra dimensions for the spherical collapse. Of particular
interest would be the relative strength of the contributions due
to the increase in dimension which favors BH and the negative
$\gamma$ which favors NS. For this, following the method of Ref. \cite{gj},
we first obtain an exact
solution of the Einstein equation in HD for a fluid with radial
pressure, satisfying the equation of state, $P_r = \gamma\rho$,
and vanishing tangential pressure. We shall then analyze the tug
of war between BH and NS. The most remarkable result that emerges
from this analysis is the interplay between the dimension $D$ of
space-time and the equation of state parameter $\gamma$. It turns
out that the final outcome of the collapse for dust in $D$,
$\gamma = 1$ in $D-1$ and  ${\gamma = -1}$ in $D+1$ dimensions is
the same. That is these three cases are indistinguishable under
gravitational collapse. This is the main result that we would like
to share through this communication.

We write, for the $D$ = $n + 2$ dimensional spherically symmetric
space-times, the metric in the comoving coordinates
\begin{equation}
ds^2 = e^{2 \psi}dt^2 - e^{2 \vartheta}dr^2 -Y^2 d \Omega^2,
\label{eq:me}
\end{equation}
where $\psi$, $\vartheta$ and $Y$ are function of $r$ and $t$, and
$d\Omega^2$ is the metric on an $n$-sphere.

The Einstein equation, $G_{ab} = -\kappa T_{ab}$ where $T^a_b =
\mbox{diag}(\rho, -P_r, -P_{\theta},.\;.\;.\;-P_{\theta})$,  after
some manipulations lead to the following system of equations:
\begin{eqnarray}
m' & = &\kappa \frac{(n-1)}{n} Y^n Y'\rho, \label{eqm1} \\
\dot{m} & = & -\kappa \frac{(n-1)}{n} Y^n \dot{Y}P_r,  \label{eqm2}\\
\psi'(\rho+P_r)& = &\left[n(P_r-P_{\theta}) \frac{Y'}{Y} +
{P_r}'\right]
\end{eqnarray}
where
\begin{eqnarray}
k(t,r)& = &1 - e^{-2 \vartheta}{Y'}^2, \label{eqk} \\
m(t,r)& = &\frac{n-1}{2}Y^{n-1} \left[e^{-2 \psi}
\dot{Y}^2+k(t,r)\right].\label{eqmf}
\end{eqnarray}
Here dot and prime stand respectively for the differentiation with
respect to $t$ and $r$. The `conservation` equation $\nabla_a
T^a_b = 0$ reads
\begin{eqnarray}
\dot{\vartheta} & = &
\frac{\dot{Y}'}{Y'} - \psi '\frac{\dot{Y}}{Y'} \label{eqc2} \\
\dot{\rho} & = & -(\rho + P_r)\left(\dot{\vartheta} +
n\frac{\dot{Y}}{Y}\right) + n(P_r - P_{\theta})\frac{\dot{Y}}{Y},
\end{eqnarray}

In this model we take zero tangential pressure and the equation of
state for the radial pressure, $P_r = \gamma \rho$ with $-1<
\gamma< 1$. Setting the tangential pressure $P_{\theta} = 0$ in
Eq.~(\ref{eqc2}), yields
\begin{equation}
\psi' (\rho+P_r) = \left[n P_r \frac{Y'}{Y} + {P_r}' \right].
\label{ptr}
\end{equation}
Further we consider the marginally bound case which means $k(t,r)
= 0$ and so we get $\dot{\vartheta} = {\dot{Y}'}/{Y'}$. Then the
above equation implies $\psi = \psi(t)$ which could be absorbed by
redefining $t$ to proper time $\tau$ via $\tau = \int e^{\psi(t)}
dt + f(r)$. We thus obtain the general solution
\begin{equation}
Y(\tau,r) = A^{\frac{1}{(n+\gamma+1)}}
r^{{(n+\gamma-1)}/{(n+\gamma+1)}} \left[ \tau_0(r) - \tau
\right]^{\frac{2}{(n+\gamma+1)}}, \label{sy}
\end{equation}
with
\begin{equation}
\rho = C \frac{r^{n-2}}{Y^n} \left(\frac{r}{Y}\right)^{1+\gamma}
\end{equation}
where $\tau_0(r) = {r}/{\sqrt A}$,
\[
A = \frac{(n+\gamma+1)^2}{4} \left (\frac{2 \kappa}{n} C \right)
\]
and $C$ is a constant. Here, we note that the solution~(\ref{sy}),
as in the 4D case, is exact only for $\gamma=-1,\;0 $ and
approximate otherwise.  For other value of $\gamma$,
$G_{\theta_1\theta_1}\neq0$ and it in fact reads as
\begin{equation}
P_{\theta}\propto (1+\gamma)\gamma r^{2(n+\gamma-1)/(n+\gamma+1)}
{\mathcal{F}}(t,r),
\end{equation}
where ${\mathcal{F}}(t,0)\propto t^{-(n+\gamma+3)/(n+\gamma+1)}$.
It vanishes as $r$ goes to zero. For $\gamma = 0,-1$, we have the
exact solution else it is approximately valid close to the central
singularity. This is precisely the region of interest as
singularity is approached. The weak energy condition which
requires $\rho \geq 0$, $(\rho + P_r) \geq 0$,
 $ (\rho + P_{\theta}) \geq 0$ is clearly satisfied.

 The apparent horizon is formed when the boundary of trapped surface is
formed at $(n-1)Y^{(n-1)} = 2 m$. The corresponding time $\tau =
\tau_{\mbox{ah}}$  is given by
\begin{equation}
\tau_{\mbox{ah}}(r) = \tau_0(r) \left[ 1 -
\Theta^{{(n+\gamma+1)}/{(n+\gamma-1)}}
 \right],
\end{equation}
with $\Theta = \sqrt{{2\kappa C}/{n}}$. As in the 4D case, it can
be shown that $\tau > \tau_{\mbox{ah}}$ for all $r>0$ and
$\tau_0(0)= \tau_{\mbox{ah}}(0)$ at $r=0$. It then follows that
only the central singularity at $r=0$ could be naked while the
others with $r > 0$ are all censored. For studying the causal
structure of the singularity, we follow the outgoing radial null
geodesics and check whether some of them meet the singularity in
the finite past.
 The equation of radial null geodesics is
\begin{equation}
\frac{d \tau}{dr} = \pm Y' =\pm \frac{Y}{r} \frac{1}{(n+\gamma
+1)} \left[n +\gamma - 1 + 2 \left(1 - \frac{\tau}{\tau_0}
\right)^{-1} \right].
  \label{rng}
\end{equation}
Along the outgoing radial null geodesics we have
\begin{equation}
\frac{dY}{dr} = Y' + \dot{Y} \left(\frac{d\tau}{dr} \right) =
Y'(1+\dot{Y}).  \label{rng1}
\end{equation}
Using the standard procedure, we introduce the auxiliary variables
$u, ~X:$
\begin{eqnarray}
u & = & r^\alpha , \;  \alpha > 0, \\
 X & = & \frac{Y}{u}.
\end{eqnarray}
In the limit of approach to the singularity we write
\begin{eqnarray}
X_0 & = & \lim_{Y \rightarrow 0,u \rightarrow 0} \frac{Y}{u} =
\lim_{Y \rightarrow 0,u \rightarrow 0} \frac{dY}{du}
 =\lim_{Y \rightarrow 0,r \rightarrow 0} \frac{1}{\alpha
r^{\alpha-1}}\frac{dY}{dr}\nonumber \\
 & = & \lim_{Y \rightarrow 0,r
\rightarrow 0} \frac{1}{\alpha r^{\alpha-1}} Y'(1+\dot{Y}).
\label{re}
\end{eqnarray}
In order to obtain the root equation, we first obtain an explicit
expression for $Y'$.  Now, from Eq.~(\ref{sy}), we have
\begin{equation}
\tau(Y,r) = \frac{r}{\sqrt A} \left[1-  \left(\frac{Y}{r}
\right)^{(n+\gamma+1)/2} \right]. \label{tr}
\end{equation}
By differentiating Eq.~(\ref{tr}) with respect to r, we obtain:
\begin{equation}
Y'(Y,r) = \frac{Y}{r} \frac{1}{(n+\gamma+1)} \left[ n + \gamma -1
+ 2 \left(\frac{r}{Y} \right)^{(n+\gamma+1)/2} \right]. \label{yd}
\end{equation}
We insert Eq.~(\ref{yd}) into (\ref{re}) to get
\begin{eqnarray}
X_0 & = & \lim_{r \rightarrow 0} \frac{X}{\alpha(n+\gamma+1)}\Big[
\left[ n + \gamma -1 + 2
\frac{r^{(1-\alpha)\times(n+\gamma+1)/2}}{X^{(n+\gamma+1)/2}}
\right]\nonumber \\
& & \times \left[ 1 - \Theta \frac{r^{(1-\alpha)(n+\gamma-1) /
2}}{X^{(n+\gamma-1)/2}} \right] \Big]. \label{re1}
\end{eqnarray}
A self consistent solution occurs for $\alpha = 1$. Therefore the
desired root equation becomes
\begin{equation}
y_0^{2(n + \gamma)} + \frac{n+\gamma-1}{2} \Theta
y_0^{(n+\gamma+1)} - y_0^{(n+\gamma-1)} + \Theta = 0 \label{ae}
\end{equation}
where $y_0 = \sqrt{X_0}$. Clearly this equation remains unaltered
so long as $n + \gamma$ remains fixed. Since $n$ can take only
integral value, $n + \gamma$ could remain fixed only if $\gamma$
takes integral value, which could only be $0, \pm1$. The equation
remains invariant for (${D-1},\;\gamma=1$), (${D},\; \gamma=0$)
and (${D+1},\; \gamma=-1$) where $n = {D-2}$.

The existence of a real positive root to this algebraic equation
is necessary and sufficient condition for the existence of NS.
The values of the roots give the tangents of the
escaping geodesics near the singularity. Thus, the occurrence of
positive roots would imply the violation of the strong CCC, though
not necessarily of the weak form. Hence in the absence of positive
real roots, the collapse will always lead to a black hole.  The
critical slope would be given by the double root, marking the
threshold between BH and NS.  It is
interesting to see that for each $D$, there exists a
$\Theta_{\mbox{crit}}^D$ such that singularities are always naked
for all $\Theta \in (0, \Theta_{\mbox{crit}}^D]$, i.e., for each
$D$ there exists a non zero measure set of $\Theta$ values giving
rise to NS and consequently violating CCC.

 From the above root equation, which is the master equation
governing the end result, BH v/s NS, follows the main result of
the paper that dust in $D$, de Sitter like in ${D-1}$ and anti de
Sitter like in ${D+1}$ dimensions are absolutely equivalent for
the end result of the collapse. (Here, by de Sitter and anti-de Sitter, we
simply mean $\gamma = \pm1$.) This is a remarkable new result
which interestingly hooks space-time dimension with the equation
of state parameter. Clearly the above root equation always has a
positive root for the $4$-dimensional AdS-like ($\gamma = -1$)
collapse and hence singularity is in this case always naked. It is
however known that increase in $D$ makes gravity stronger and
thereby it favors BH against NS indicated by the shrinkage of the
NS producing parameter window in the initial data set. Positive
pressure also has similar contribution while negative pressure has
the opposite effect. That is why for $\gamma=-1$, as $D$ increases
both BH/NS could occur while it is all NS for $D=4$. That is
strengthening of gravity is stronger due to increase in $D$ than
the opposite negative pressure contribution. The parameter window
in the initial data set leading to NS shrinks. Though here we have
essentially considered radial pressure, the result would be valid
in general for any gravitational collapse with pressure.

\noindent SGG and DWD would like to thank IUCAA, Pune for
hospitality while this work was done.  SGG also thanks to
Gridhar Vishwanathan for providing him support in computation.


\begin{references}
\bibitem{rp} R.~Penrose, {Riv del Nuovo Cimento} {\bf 1}, 252
 (1969); in {\it General Relativity}, {\it an Einstein Centenary
 Volume}, edited by S. W.~Hawking and W.~Israel (Cambridge
 University Press, Cambridge, England, 1979).
\bibitem{r1} P.S.~Joshi, {\it Global Aspects in Gravitation and
Cosmology} (Clendron Press, Oxford, 1993); C.J.S.~Clarke, {
Class. Quantum Grav.} {\bf 10}, 1375 (1993); T.P.~Singh, {J.
Astrophys. Astron.} {\bf 20}, 221 (1999); P.S.~Joshi, {Pramana}
{\it 55}, 529 (2000).
\bibitem{rm} R.M.~Wald, {\it Gravitational Collapse and Cosmic
Censorship}, gr-qc/9710068.
\bibitem{kt} K.S.~Thorne, in {\it Magic without magic}, edited by
J.R.~Klander (Freeman, San Francisco, 1972).
\bibitem{he} S.W.~Hawking and G.F.R.~Ellis, {\emph{The Large Scale
Structure of Space-Time}}  (Cambridge; Cambridge University Press)
(1973).
\bibitem{dc} D.M.~Eardley and L.~Smarr, {Phys. Rev. D}
{\bf 19}, 2239 (1979); D.M.~Eardley, in {\it Gravitation in
Astrophysics}, edited by B.~Carter and J.B.~Hartle (NATO Advanced
Study Institute, Series B: Physics, Vol. 156), (Plenum Press, New
York, 1986) pp 229-235; B.~Waugh and K.~Lake, {Phys. Rev. D} {\bf
38}, 1315 (1988); J.P.S.~Lemos, {Phys. Lett. A} {\bf 158}, 271
(1991);  J.P.S.~Lemos, {Phys. Rev. Lett.} {\bf 68}, 1447 (1992);
I.H.~Dwivedi and P.S.~Joshi, {Class. Quantum Grav.} {\bf 9}, L69
(1992); P.S.~Joshi and I.H.~Dwivedi, {Phys. Rev. D} {\bf 47}, 5357
(1993); P.S.~Joshi and T.P.~Singh,  {Phys. Rev. D} {\bf 51}, 6778
(1995);  S. Jhingan, P.S.~Joshi, T.P.~Singh, Classical and Quantum
Gravity, {\bf 13}, 3057(1996); S. Jhingan and P.S.~Joshi,  Annals
of the Israel Physical Society {\bf 13}, 357 (1998).
\bibitem{pf}  A.~Ori and T.~Piran, {Phys. Rev. Lett.} {\bf 59},
2137 (1987); A.~Ori and T.~Piran, {Phys. Rev. D} {\bf 42}, 1068
(1990); P.S.~Joshi and I.H.~Dwivedi, {Commun. Math. Phys.} {\bf
146}, 333 (1992); P.S.~Joshi and I.H.~Dwivedi, Lett. Math. Phys.
{\bf 27}, 235 (1993); F.I.~Cooperstock, S.~Jhingan, P.S.~Joshi and
T.P.~Singh, {Class. Quantum Grav.} {\bf 14}, 2195 (1997);
T.~Harada, {Phys. Rev. D} {\bf 58}, 104015 (1988); G.~Magli,
{Class. Quantum Grav.} {\bf 14}, 1937 (1997); {\bf 15} 3215
(1998); T.~Harada, H.~Iguchi and K.I.~Nakao, {Phys. Rev. D} {\bf
58}, 041502 (1998); J.F.V.~Rocha, A.~Wang and N.O.~Santos, {Phys.
Lett. A} {\bf 255}, 213 (1999); S.~Jhingan and G.~Magli, {Phys.
Rev. D} {\bf 61}, 124006 (2000); S.M.C.V.~Gon\c{c}alves, S.
Jhingan and G. Magli, Phys. Rev. D Vol. 65, 064011 (2002).
\bibitem{gj} S.M.C.V.~Gon\c{c}laves and S.~Jhingan.
{Gen. Rel. Grav.} {\bf 33}, 2125 (2001).
\bibitem{rs} L.~Randall and R.~Sundrum, {Phys. Rev. Lett.}
{\bf 83}, 3370 (1999); {\bf 83}, 4690 (1999).
\bibitem{hvc} S.G.~Ghosh and R.V.~Saraykar, {Phys. Rev. D} {\bf 62},
107502 (2000); S.G.~Ghosh and N.~Dadhich, {Phys. Rev. D} {\bf 64},
047501 (2001).
 \bibitem{hdc} A.~Banerjee, A.~Sil and S.~Chatterjee {Astrophys. J.}
 {\bf 422}, 681 (1994); A.~Sil and S.~Chatterjee {Gen. Relativ. Gravit.}
 {\bf 26}, 999 (1994); S.G.~Ghosh and A.~Beesham, {Phys. Rev. D} {\bf 64},
124005 (2001); S.G.~Ghosh and A.~Banerjee, {Int. J. Mod. Phys. D}
{\bf 12}, 639 (2003);
\bibitem{htt}S.G.~Ghosh and N.~Dadhich, {Phys. Rev. D} {\bf 65},
127502 (2002); H.~Kim, S.~Moon and J.~Yoe, {JHEP} {\bf 0202:046};
(2002). A.V.~Frolov {Class. Quantum Grav.} {\bf 16}, 407 (1999).
\bibitem{hdo} J.~Soda and K.~Hirata, {Phys. Lett. B} {\bf 387}, 271 (1996);
J.F. Villas da Rocha, Anzhong Wang, Class. Quantum Grav. {\bf 17},
2589(2000); Jaime F. Villas da Rocha, Anzhong Wang, {\it
Gravitational Collapse of Perfect Fluid in N-Dimensional
Spherically Symmetric Spacetimes,} gr-qc/9910109; S.G.~Ghosh and
D.W.~Deshkar, {Int. J. Mod. Phys. D} {\bf 12},  (2003).
\bibitem{hos} A.~llha and J.P.S.~Lemos,  {Phys. Rev. D} {\bf 55}, 1788
(1997);   A.~llha, A.~Kleber and J.P.S.~Lemos, {J. Math. Phys.} {\bf 40},
 3509 (1999).
\bibitem{gj1} R.~Goswami and P.S.~Joshi "{\it Is Cosmic Censorship
valid in Higher Dimensions}", gr-qc/0212097.
\end{references}
\end{document}